\documentclass[draftclsnofoot,onecolumn,12pt]{IEEEtran}
\usepackage{cite}
\usepackage{graphicx,color}
\usepackage{float}
\ifCLASSINFOpdf
\else
\fi
\usepackage[cmex10]{amsmath}

\usepackage{bm}
\usepackage{amssymb}
\usepackage{algorithm}
\usepackage{algorithmic}
\usepackage{array}
\usepackage{fixltx2e}

\usepackage{indentfirst}
\usepackage{booktabs}
\usepackage{multirow}
\usepackage{bigstrut}
\usepackage{setspace}
\usepackage{booktabs}
\usepackage{xcolor}
\usepackage{amsthm}

\IEEEoverridecommandlockouts

\newtheorem{Theo}{Theorem}
\newtheorem{Lem}{Lemma}

\newtheorem{Def}{Definition}

\begin{document}
%
\title{Energy-Efficient Data Transmission with Non-FIFO Packets}

\author{\IEEEauthorblockN{Qing Zhou, Nan Liu}\\
\IEEEauthorblockA{National Mobile Communications Research Laboratory, Southeast University, Nanjing 210096, China\\
Email:\{qingzhou and nanliu\}@seu.edu.cn}
\thanks{
This work is partially
supported by the National Basic Research Program of China
(973 Program 2012CB316004), the National Natural Science Foundation of China
under Grants $61201170$, $61271208$ and $61221002$ and Qing Lan Project.}
}

\maketitle

\begin{abstract}
This paper investigates the problem of energy-efficient packet transmission with arbitrary arrival instants and deadline constraints over a point-to-point Additive White Gaussian Noise (AWGN) channel. This is different from previous work where it is assumed that the packets follow a First-In-First-Out (FIFO) order in that the packets that arrive earlier will have a deadline that is also earlier.
We first investigate the necessary and sufficient conditions of the optimal transmission scheduler. We then propose an algorithm which finds the transmission schedule of each packet in the order of the packets with the largest transmission rate to the packets with the smallest transmission rate. Finally, we show that our algorithm satisfies the sufficient conditions of the optimal transmission scheduler and thus, is optimal.
\end{abstract}

\IEEEpeerreviewmaketitle

\section{Introduction}
Energy efficiency (EE) is an emerging issue for designing new communication systems to achieve significant energy savings, which will cut the operational costs as well as the  emission of carbon dioxide. References \cite{Berry2002},\cite{Uysal-Biyikoglu2002} showed that transmitting data flow with a low constant rate is an efficient method to reduce energy expenditure due to the fact that the transmission power is an increasing and strictly convex function of transmission rate.
However, most of the current data services such as Voice over Internet Phone (VoIP) and video conference are often time-critical and delay-sensitive, therefore the Quality of Service (QoS) is an important factor which should be considered when we designing energy efficient realtime communication systems.

To this end, there have been many strategies put forth to address the energy-efficient transmission problems  \cite{Uysal-Biyikoglu2002,Zafer2009,Chen2008}. In \cite{Uysal-Biyikoglu2002}, the authors considered a transmission energy minimization problem for packet transmission with a single deadline constraint over a point-to-point AWGN time-invariant channel. A ``lazy scheduler'' was proposed as the optimal transmission strategy to achieve energy efficient packet transmission under the causality and deadline constraints. Generalizing \cite{Uysal-Biyikoglu2002} with respect to deadline constraints, \cite{Zafer2009,Chen2008} studied similar problems under individual deadline constraints: \cite{Zafer2009} posed the problem as a continuous time optimization and proposed a calculus approach to obtain the ``optimal departure curve'', which had a simple and appealing graphical visualization, which was named ``string tautening'' in \cite{Nan2014a}; in \cite{Chen2008}, a recursive optimal scheduling algorithm was put forward to find out the optimal policy to realize minimal energy consumption. In addition, \cite{Nan2014a,Jin2014} takes the circuit power consumption into consideration and investigated energy efficient transmissions of bursty data packet with individual deadlines under non-ideal circuit power consumption. In another relevant research field of energy harvesting, \cite{Xu2014,Bai2011,Ozel2011,Yang2012} study the throughput maximizing problem or transmission time minimization problem for packet transmission subject to the causality constraint of energy arrivals and packet arrivals as well as the capacity constraint of the battery.

All the works in \cite{Uysal-Biyikoglu2002,Zafer2009,Chen2008,Nan2014a,Jin2014,Xu2014,Bai2011,Ozel2011,Yang2012} assumed that all the packets are \emph{FIFO packets}, i.e.,  the individual deadlines of the data flow were consistent with the order of their arrival instants. However, in practical wireless communication systems, different applications and services have different requirements for packet delay, e.g., real-time voice or video and real-time games have high requirements on packet delay; while, buffered video streaming and TCP based services, such as www, ftp and e-mail, are less strict in terms of delay.
Therefore, it is very much possible that a packet that has arrived later must depart before a packet that arrived earlier. In other words, the consistency of the order of the deadlines and the arrival instants does not always hold.

Thus, in this paper, we investigate the problem of energy-efficient packet transmission with arbitrary arrival instants and deadline constraints over a point-to-point Additive White Gaussian Noise (AWGN) channel.
We first derive the necessary and sufficient conditions of the optimal transmission scheduler. We then propose an algorithm which finds the transmission schedule of each packet in the order of the packets with the largest transmission rate to the packets with the smallest transmission rate. Finally, we show that our algorithm satisfies the sufficient conditions of the optimal transmission scheduler and thus, is optimal.

\section{system model}
\subsection{Data Flow Model}
In this paper, we consider a point-to-point wireless link over an AWGN channel which is assumed to be time-invariant. There are $N$ packets randomly arriving at the transmitter buffer in sequence, and the set of the packets is denoted as $\mathcal{P} = \left\{ {{P_1},{P_2}, \ldots ,{P_N}} \right\}$. The key attributes of each packet can be expressed as ${I_i} = \left( {{B_i},{t_{a,i}},{t_{d,i}}} \right)$, $1 \le i \le N$, where ${B_i}$ is the size of the $i$-th packet, and ${t_{a,i}}$ and ${t_{d,i}}\left( { > {t_{a,i}}} \right)$ represent the corresponding arrival instant and the deadline of Packet $i$, respectively.
For the offline transmission scheme, we assume that the key attributes of each packet as well as the channel state information (CSI) are a priori known at the transmitter, which is the assumption also made in \cite{Uysal-Biyikoglu2002,Zafer2009,Chen2008,Nan2014a}. For the online transmission scheme, we assume that the key attributes of each packet is known causally.

Without loss of generality, the first packet is assumed to arrive at instant 0, and the packets arrived in sequence, i.e., $0 = {t_{a,1}} < {t_{a,2}} <  \cdots  < {t_{a,N}}$. Previous works  \cite{Uysal-Biyikoglu2002,Zafer2009,Chen2008,Nan2014a,Jin2014} assumed that the deadlines of the packets follow the same order as the arrival times in the sense that ${t_{d,1}} < t_{d,2}<  \cdots  < t_{d,N}$. In this work, we consider a generalized scenario with respect to the deadline constraints, i.e., the deadlines of the packets are arbitrary. Hence, the condition ${t_{d,1}} < t_{d,2}<  \cdots  < t_{d,N}$ assumed by previous work does not hold. For a packet $P_i$, $i \in \{2,\cdots,N\}$, if it satisfies $t_{a,k}<t_{a,i}<t_{d,i}<t_{d,k}$, for some $k<i$, then we call packet $P_i$ a {\it{non-FIFO packet}}.
We remove the repeated instants in $\begin{bmatrix} t_{a,1},\cdots, t_{a,N}, t_{d,1}, \cdots, t_{d,N} \end{bmatrix}$ and arrange them in ascending order, this is denoted as the set of ascending instants
$\Gamma = \{t_0=0, t_1, \cdots, t_M=T\}$, where $T = \max \left\{ {{t_{d,i}}|1 \le i \le N} \right\}$ and $M$ is the number of time instants left after removing the repeated time instants.

Next, we provide some definitions based on the set of ascending instants $\Gamma =\{t_0=0, t_1, \cdots, t_M=T\}$: 
\begin{Def}
An {\it{epoch}} is defined as the interval of two adjacent instants, i.e., $\mathcal{E}_j = [t_{j-1},t_j], j=1,2,\cdots,M$, and the length of epoch $\mathcal{E}_j$ is denoted as $|\mathcal{E}_j|$, where $|\mathcal{E}_j|=t_j - t_{j-1},j=1,2,\cdots,M$.
\end{Def}
\begin{Def}
The {\it{life time duration}} $\mathcal{L}_i$ of Packet $P_i$ is defined as the time interval between the arrival instant and deadline of Packet $P_i$, i.e., $\mathcal{L}_i = [t_{a,i},t_{d,i}], i=1, 2, \cdots, N$, and the length of $\mathcal{L}_i$ is denoted as $|\mathcal{L}_i|$, where $|\mathcal{L}_i|=t_{d,i}-t_{a,i}, i=1,2, \cdots, N$.
\end{Def}

\begin{Def}
Denote by $\mathcal{C}_i$ as the set of epochs which are contained in Packet $P_i$'s life time duration $\mathcal{L}_i$, i.e., $\mathcal{C}_i=\{j|\mathcal{E}_j \subseteq \mathcal{L}_i\}$. Further denote by $\mathcal{F}_j$ as the set of packets  are can be transmitted in epoch $\mathcal{E}_j$, i.e., $\mathcal{F}_j=\{i|\mathcal{E}_j \subseteq \mathcal{L}_i\}$, where $i=1, 2, \cdots, N$ and $j= 1, 2, \cdots, M$.
\end{Def}

\subsection{Transmission Model}
We let $p\left( t \right)$ signify the transmission power at time $t$ when the transmission rate is $r\left( t \right)$. The relationship between $p\left( t \right)$ and $r\left( t \right)$ can be described using the function $f$ as:
\begin{align}\label{PowerRate}
 p(t) = f(r(t))
\end{align}
where $f\left( \cdot \right)$ is a convex and increasing function defined on $[0, \infty]$. In addition, $p\left( x \right) \geq 0$ for all $t \in [0,\infty]$.

Shannon's capacity formula over an AWGN channel provides a typical example for the function $f$ as follows:
\begin{align}\label{Shannon}
r(t) = \frac{1}{2}\log\left(1+\frac{p(t)}{N}\right)
\end{align}
where $N$ is the variance of the channel noise. We may rewrite equation (\ref{Shannon}) as $p\left( t \right) = N\left( {{2^{2r\left( t \right)}} - 1} \right)$. It can be easily verified that the expended power is  a convex and increasing function of the transmission rate. More examples of the function $f$ is provided in \cite{Uysal-Biyikoglu2002}.

\subsection{Problem Formulation}
The problem of finding the optimal transmission strategy of the packets to minimize the transmission energy can be formulated as follows:
\begin{subequations}\label{ProblemFormulation}
  \begin{align}
  \min_{r(t)}~~~&E\left(r(t) \right) \triangleq \int_0^T {f\left(r(t) \right)} dt \\
  \mbox{subject~to}~~&\int_{t_{a,i}}^{t_{d,i}} {r(t)\Delta\left(P(t),P(i) \right)} dt = B_i,~~i \in [1,N].
  \end{align}
\end{subequations}
where $P(t)$ is the packet which is being transmitted at instant $t$, $\Delta(a,b)$ is the indicate function such that
\begin{equation}
\Delta (a,b)  = \left\{ {\begin{array}{*{20}{c}}
{1,}&{a = b;}\\
{0,}&{a \ne b.}
\end{array}} \right.
\end{equation}
Note that the constraint in (\ref{ProblemFormulation}b) implies that the scheduler must satisfy the causality constraint, i.e., no packet data can be transmitted before it has arrived, and the deadline constraint, i.e., we must finish transmitting all of the packet's data before its deadline. We call a scheduler that satisfies the causality constraint and the deadline constraint \emph{a feasible scheduler}.

Based on the convexity of the function $f(\cdot)$, we have the following lemma.

\begin{Lem} \label{ConstantRate}
In the optimal transmission schedule, each packet should be transmitted at a constant rate.
\end{Lem}

Proof: The proof follows from \cite[Theorem 1]{Zafer2009}. This result is true due to the convexity of the function $f(\cdot)$.               \hfill $\Box$

According to Lemma \ref{ConstantRate}, each packet should be transmitted at a constant rate. Denote the constant transmission rate of packet $P_i$ as $r_i$, and the set of transmission rates of all the packets by ${\bf{r}}=\{r_1, r_2, \cdots, r_N\}$.
Further denote by $\tau_{i,j}, i=1,2,\cdots,N, j=1,2,\cdots,M$ as the transmission time of packet $P_i$ in Epoch $\mathcal{E}_j$. Note that Packet $P_i$ can only be transmitted in epochs that are contained in its life time, hence we have $\tau_{i,j}=0$, for $j \notin \mathcal{C}_i$. Denote the set of all $\tau_{i,j}$ as $\bm{\tau}=\{\tau_{i,j}, i=1,2,\cdots,N, j=1,2,\cdots,M\}$. Therefore, the original problem (\ref{ProblemFormulation}) can be equivalently reformulated as follows:
\begin{subequations}\label{OptimizationPro}
  \begin{align}
  \min_{\mathbf{r},\bm{\tau}}~&\sum_{i=1}^N \frac{B_i}{r_i}f(r_i)\\
  \mbox{subject~to}~&\frac{B_i}{r_i} - \sum_{j \in \mathcal{C}_i} \tau_{i,j} = 0,~&1 \le i \le N; \\
  ~&\sum_{i \in \mathcal{F}_j} \tau_{i,j} -|\mathcal{E}_j| \le 0,~& 1 \le j \le M;\\
  ~&\tau_{i,j} \ge 0,~&1 \le j \le M,i \in \mathcal{F}_j;\\
  ~& r_i \ge 0,~&1 \le i \le N.
  \end{align}
\end{subequations}
where constraint (\ref{OptimizationPro}b) denotes that the sum of the transmission times of Packet $P_i$ in each epoch should be equal to the total time of transmission of Packet $P_i$, which is equal to $\frac{B_i}{r_i}$, constraint (\ref{OptimizationPro}c) indicates that the sum of the transmission times of all the packets feasible in epoch $\mathcal{E}_j$ should not exceed the length of the epoch $|\mathcal{E}_j|$, constraints (\ref{OptimizationPro}d) and (\ref{OptimizationPro}e) mean that the transmit-rate of packet $P_i$, $i = 1,\cdots,N$ as well as the transmission time in its feasible epoch cannot be negative.

\begin{Lem}\label{NonIdlingtrans}
In the optimal transmission schedule, the transmission must be ``non-idling'' in each epoch $\mathcal{E}_j$, i.e., ${\sum\limits_{i \in \mathcal{F}_j } {{\tau _{i,j}^*} - \left| {{\mathcal{E}_j}} \right|}  = 0}$, $j=1,\cdots,M$.
\end{Lem}

Proof:  The proof follows from \cite[Section III]{Uysal-Biyikoglu2002}. This result is true due to the monotonicity of the function $f(\cdot)$.

\section{The Optimal Off-line Policy}
Since the objective function is convex and all the constraints are linear, problem (\ref{OptimizationPro}) is a standard convex problem. Any convex programming tools such as the gradient-type (or interior-point) iterative primal dual algorithms, can be employed to solve this problem. However, these general algorithms have high complexity, e.g., the computation complexity of interior-point method is approximately $\mathcal{O}(N)^{3.5}$ and that of ellipsoid method is $\mathcal{O}(N)^{6}$ \cite{Boyd2004,ye2011interior}, and cannot yield the specific structure of the optimal policy. Hence, we will develop a lower complexity and more insightful scheduler for the optimization problem  in (\ref{OptimizationPro}).

\subsection{Optimality Conditions}
We first derive the KKT optimality conditions  of problem (\ref{OptimizationPro}), let $\bm{\Xi}=\{\lambda_i, \beta_j, \gamma_{i,j}, \eta_i\}$, where $\lambda_i, i=1,\cdots,N$, $\beta_j, j=1,\cdots,M$, $\gamma_{i,j}$, $j=1,\cdots,M, i \in \mathcal{F}_ j$ and $\eta_i, i=1,\cdots,N$ denote the lagrange multipliers associated with the constraints in (\ref{OptimizationPro}b)-(\ref{OptimizationPro}e), respectively. Hence, the Lanrangian function of (\ref{OptimizationPro}) for any $\beta_j \ge0$, $\gamma_{i,j} \ge0$ and $\eta_i \ge0$ can be expressed as:
\begin{align}
L({\bf{r}, \bm{\tau}, \bm{\Xi}}) &= \sum_{i=1}^N \frac{B_i}{r_i}f(r_i) +\sum_{i=1}^N \lambda _i \left( \frac{B_i}{r_i}-\sum_{j \in \mathcal{C}_i} \tau_{i,j} \right)\nonumber\\
&+ \sum_{j=1}^M \beta _j \left(\sum_{i \in \mathcal{F}_j} \tau_{i,j} -|\mathcal{E}_j| \right) -\sum_{j=1}^M \sum_{i \in \mathcal{F}_j} \gamma_{i,j}\tau_{i,j}-\sum_{i=1}^N \eta_i r_i\\
&= \sum_{j=1}^M \sum_{i \in \mathcal{F}_j}(-\lambda _i+\beta _j-\gamma_{i,j})\tau_{i,j}\nonumber\\
&+ \sum_{i=1}^N \left( \frac{B_i}{r_i}f(r_i) + \lambda _i\frac{B_i}{r_i}-\eta_i r_i\right)+C(\bm{\Xi}) \label{Nan01}
\end{align}
where (\ref{Nan01}) follows from the fact that $\sum\limits_{i=1}^N \sum\limits_{j \in \mathcal{C}_i} \tau_{i,j}=\sum\limits_{j=1}^M \sum\limits_{i \in \mathcal{F}_j} \tau_{i,j}$, and  $C\left(\bm{ \Xi}  \right) \triangleq \sum\limits_{j = 1}^M {{\beta _j}\left| {{\mathcal{E}_j}} \right|} $, which is a term independent of $\bf{r}$ and $\bm{\tau}$. Let $({\bf{r^*}},{\bm{\tau^*}})$ represent the optimal solution of the problem (\ref{OptimizationPro}) and $\bm{\Xi^*}=\{\lambda_i^*, \beta_j^*, \gamma_{i,j}^*, \eta_i^*\}$ denote the optimal Lagrange multiplier vector for its dual problem. The KKT conditions can be obtained by taking the derivatives of $L( {{\bf{r}},\bm{\tau} ,\bm{\Xi} })$ with respect to $r_i^*$ and $\tau_{i,j}^*$ as:
\begin{subequations}\label{KKT_Conditions}
\begin{align}
~&-\frac{B_if(r_i^*)}{{r_i^*}^2}+\frac{B_if'(r_i^*)}{r_i^*}-\frac{B_i \lambda_i^*}{{r_i^*}^2}-\eta_i^*=0,~1\le i \le N;\\
~&- {\lambda _i^*} + {\beta _j^*} - {\gamma _{i,j}^*} = 0,~1 \le i \le N,j \in \mathcal{C}_i.
\end{align}
\end{subequations}
where $f'(\cdot)$ is the derivative of $f(\cdot)$, which is positive and monotonically increasing function since $f(\cdot)$ is increasing and strictly convex function.

Furthermore, the optimal non-negative Lagrangian multipliers $\beta_j ^*$, $\gamma_{i,j}^*$ and $\eta_i^*$ must satisfy the complementary slackness conditions \cite{Boyd2004}:
\begin{subequations}\label{ComplementarySlackness}
  \begin{align}
~&\beta _j^*\left( \sum_{i \in \mathcal{F}_j} \tau_{i,j}^* -|\mathcal{E}_j|  \right) = 0,~&1 \le j \le M;\\
~&\gamma_{i,j}^* \tau _{i,j}^* = 0,~&1 \le j \le M,i \in \mathcal{F}_j;\\
~&\eta _i^* r_i^* = 0, ~&{1 \le i \le N.}
  \end{align}
\end{subequations}
since $r_i^*>0$ always hold, then $\eta_i ^*= 0$ by the complementary slackness condition in (\ref{ComplementarySlackness}c) and hence (\ref{KKT_Conditions}a) can be rewritten as:
\begin{align}\label{KKT_Rewritten}
-f(r_i^*) + r_i^*f'(r_i^*)-\lambda_i^*=0,~1\le i \le N.
\end{align}
We denote $g(r_i^*) = {r_i^*}f'\left( {{r_i^*}} \right) - f\left( {{r_i^*}} \right)$ which is monotonically increasing function since $g'(r_i^*) = r_i^*f''(r_i^*) \geq 0$, where $f''(\cdot)$ is the second derivative of function $f(\cdot)$. Let $g^{-1}(\cdot)$ denote the inverse function of $g(\cdot)$, which is also a monotonically increasing function due to the monotonicity of $g(\cdot)$. The optimal transmission rate $r_i^*$ can be derived from (\ref{KKT_Conditions}b) and (\ref{KKT_Rewritten}):
\begin{align}\label{Optimal_R}
  r_i^* = g^{-1}(\beta_j^* - \gamma_{i,j}^*),~~1\le i \le N,~j \in \mathcal{C}_i.
\end{align}

We obtain the following lemma which follows from the optimality conditions (\ref{KKT_Conditions}), (\ref{ComplementarySlackness}), (\ref{KKT_Rewritten}) and (\ref{Optimal_R}):
\begin{Lem} \label{OptRateRelationship}
Consider epoch $\mathcal{E}_j$, $j=1,2,\cdots,M$, and the set of packets feasible in epoch $\mathcal{E}_j$, i.e., Packets $P_i$, where $i \in \mathcal{F}_j$. These packets are divided into 2 sets: $\Psi_j = \{i|i \in \mathcal {F}_j, \tau_{i,j}^* >0\}$, i.e., the set of packets in $\mathcal{F}_j$ that get positive transmission time in Epoch $j$ and $\bar{\Psi}_j = \{i|i \in \mathcal {F}_j,\tau_{i,j}^* =0\}$, i.e., the set of packets in $\mathcal{F}_j$ that get zero transmission time in Epoch $j$. For the optimal transmission scheduler, the following must hold:
\begin{itemize}
\item[(1).] The transmission rates $r_i$ for $i \in \Psi_j$ are all equal.
\item[(2).] $r_i \geq r_k$, $\forall i \in \Psi_j, \forall k \in \bar{\Psi}_j$.
\end{itemize}
%
%
\end{Lem}

Proof: Please see Appendix \ref{Proof_OptRateRelationship}. \hfill $\Box$

Lemma \ref{OptRateRelationship} says that for the packets with positive transmission time in epoch $\mathcal{E}_j$, their transmission rate is the same. For packets that are in $\mathcal{F}_j$ but with no transmission time in Epoch $j$, their transmission rate can not be larger than that of the packets with positive transmission time in Epoch $j$.

%

All the properties presented in Lemmas \ref{ConstantRate},  \ref{NonIdlingtrans} and \ref{OptRateRelationship} are necessary conditions for the optimal transmission schedule. Next, we will show that all the properties in Lemmas \ref{ConstantRate},  \ref{NonIdlingtrans} and \ref{OptRateRelationship} are also sufficient conditions for optimality.

\begin{Theo}\label{NS Conditions}
If a feasible scheduler satisfies Lemmas \ref{ConstantRate},  \ref{NonIdlingtrans} and \ref{OptRateRelationship}, then it is the optimal scheduler.
\end{Theo}

Proof: Please see Appendix \ref{Proof_NS Conditions}.\hfill $\Box$

\subsection{The Optimal Transmission Scheduler}

Although the conditions in Lemmas \ref{ConstantRate},  \ref{NonIdlingtrans} and \ref{OptRateRelationship} are necessary and sufficient conditions for the optimal transmission schedule,  they do not provide us with the optimal scheduler explicitly. We propose a scheduler in this subsection and show that it is feasible and further satisfies Lemmas \ref{ConstantRate},  \ref{NonIdlingtrans} and \ref{OptRateRelationship}, thus proving that it is an optimal scheduler.

Before we proceed, we first give a definition about a sub-interval, which is rigorously described as follows:

\begin{Def} \label{NanDef}
We define a sub-interval by $\mathcal{T}_{k,l}=[t_{a,k}, t_{d,l}]$, $k,l \in \{1,\cdots,N\}$ which contains at least one packet's life time duration, i.e., there exists a packet $\bar{k} \in \{1,2,\cdots,N\}$ such that $\mathcal{L}_{\bar{k}} \subseteq \mathcal{T}_{k,l}$. We also define the set of packets whose life time is contained in sub-interval $\mathcal{T}_{k,l}$ as $\mathcal{H}(\mathcal{T}_{k,l})$, i.e., $\mathcal{H}(\mathcal{T}_{k,l})=\{i|\mathcal{L}_i \subseteq \mathcal{T}_{k,l}\}$.
\end{Def}

Note that the start of the sub-inteval is $t_{a,k}$, i.e., packet $P_k$'s arrival instant and the end of the sub-interval is $t_{d,l}$, i.e., the deadline of packet $P_l$, where $t_{d,l}>t_{a,k}$. The length of the sub-interval is $|\mathcal{T}_{k,l}|=t_{d,l}-t_{a,k}$.
\begin{Def}
The transmission rate of the sub-interval $\mathcal{T}_{k,l}$ is defined as
\begin{align}
r(\mathcal{T}_{k,l})=\frac{\sum\limits_{i \in \mathcal{H}(\mathcal{T}_{k,l})}{B_i}}{|\mathcal{T}_{k,l}|}, \label{NanRate}
\end{align}
\end{Def}
Note that this is the minimum transmission rate of the sub-interval $\mathcal{T}_{k,l}$, since to meet the deadline constraints, all packets whose life time duration is inside sub-interval $\mathcal{T}_{k,l}$ must be transmitted inside $\mathcal{T}_{k,l}$.


Based on the Definition \ref{NanDef}, we propose the following transmission scheduler and prove its optimality. The first part of the algorithm finds the transmission rate of the packets of the optimal scheduler and the second part of the algorithm illustrate the actual transmission strategy according to the optimal transmission rate found.

\begin{algorithm}[htb]
\caption{\itshape The Optimal Scheduler}
\begin{algorithmic}[1]\label{Opt_Iter_Algorithm}
\STATE Set $\mathcal{K}=\phi$ and $\mathcal{N} = \{1,\cdots,N\}$; $\bar{t}_{a,i}=t_{a,i}$, $\bar{t}_{d,i}=t_{d,i}$, $\bar{\mathcal{L}}_i=\mathcal{L}_i$, $\forall i \in \mathcal{N}$;
\STATE Find all sub-intervals $\bar{\mathcal{T}}_{k,l}$ and  according to $(\bar{t}_{a,i}, \bar{t}_{d,l})$ for all $k,l \in \mathcal{N} \backslash \mathcal{K}$, and compute $r( \bar{\mathcal{T}}_{k,l})$ according to (\ref{NanRate})
for each sub-interval;
 \label{NanState}
\STATE Find $\bar{\mathcal{T}}_{k,l}^*$ such that $ \bar{\mathcal{T}}_{k,l}^* = \arg \max\limits_{k,l \in \mathcal{N}  \backslash \mathcal{K}} {\kern 1pt} {\kern 1pt} r( \bar{\mathcal{T}}_{k,l})$
\label{NanState02}
\STATE The transmission schedule for Packet $i \in \mathcal{H}(\bar{\mathcal{T}}_{k,l}^*)$ is the following: let $\mathcal{T}_{k,l}^*$ be $\bar{\mathcal{T}}_{k,l}^*$ shifted back to real time by performing inverses of the shifts, which were performed in previous iterations using (\ref{UpdateNan1}) and (\ref{UpdateNan2}). At any given time $t \in \mathcal{T}_{k,l}^*$, find $\bar{k}=\arg \min\limits_{t_{d,k} \geq t} t_{d,k}$, where the minimum is over all packets in $\mathcal{H}(\bar{\mathcal{T}}_{k,l}^*)$ that have arrived but have not finished transmission at time $t$, and transmit Packet $\bar{k}$ at rate $r(\bar{\mathcal{T}}_{k,l}^*)$. If no such packet can be found, remain idle at time $t$. \label{NanState04}
\STATE Update $\mathcal{K}= \mathcal{H}(\bar{\mathcal{T}}_{k,l}^*) \cup \mathcal{K}$
\STATE If $\mathcal{K}= \mathcal{N}$, then End. \label{NanState03}
\STATE Else, for $i \in \mathcal{N} \backslash \mathcal{K}$, let
\begin{align} \label{UpdateNan1}
\bar{t}_{a,i} = \left\{ {\begin{array}{*{20}{c}}
   {{\bar{t}_{a,i}},} & {{\bar{t}_{a,i}} \le {\bar{t}_{a,k}^*};}  \\
   {{\bar{t}_{a,k}^*},} & {{\bar{t}_{a,k}^*} < {\bar{t}_{a,i}} \le {\bar{t}_{d,l}^*};}  \\
   {{\bar{t}_{a,i}} - \left( {{\bar{t}_{d,l}^*} - {\bar{t}_{a,k}^*}} \right),} & {{\bar{t}_{a,i}} > {\bar{t}_{d,l}^*}.}  \\
\end{array}} \right.
\end{align}
and
\begin{align} \label{UpdateNan2}
\bar{t}_{d,i} = \left\{ {\begin{array}{*{20}{c}}
   {{\bar{t}_{d,i}},} & {{\bar{t}_{d,i}} \le {\bar{t}_{a,k}^*};}  \\
   {{\bar{t}_{a,k}^*},} & {{\bar{t}_{a,k}^*} < {\bar{t}_{d,i}} \le {\bar{t}_{d,l}^*};}  \\
   {{\bar{t}_{d,i}} - \left( {{\bar{t}_{d,l}^*} - {\bar{t}_{a,k}^*}} \right),} & {{\bar{t}_{d,i}} > {\bar{t}_{d,l}^*},}  \\
\end{array}} \right.
\end{align}
update $\bar{\mathcal{L}}_i$ according to $(\bar{t}_{a,i}, \bar{t}_{d,i})$ for all $i \in \mathcal{N} \backslash \mathcal{K}$, and go to step \ref{NanState}.
\end{algorithmic}
\end{algorithm}


The idea of the algorithm is as follows: $\mathcal{K}$ denotes the set of packets whose rate and transmission intervals have been determined, and $\bar{t}_{a,i}, \bar{t}_{d,i}$, and $\bar{\mathcal{L}}_i$ denotes the updated arrival instant, deadline constraint and life time of packet $i$ at the current iteration, respectively. At each round of iteration, find all sub-intervals $\bar{\mathcal{T}}_{k,l}$ that contain at least the life duration of one packet by testing the updated arrival instants and deadline instants of all packets whose rate has not been determined, i.e., packets who are in $\mathcal{N} \backslash \mathcal{K}$. Compute $r(\bar{\mathcal{T}}_{k,l})$ according to (\ref{NanRate}), and find the maximum  $r(\bar{\mathcal{T}}_{k,l})$ over all sub-intervals with $k,l \in \mathcal{N} \backslash \mathcal{K}$, denoted as $\bar{\mathcal{T}}_{k,l}^*$. As a consequence, the transmission rate and schedule of all packets whose updated life time duration is contained in $\bar{\mathcal{T}}_{k,l}^*$, i.e., packets in $\mathcal{H}(\bar{\mathcal{T}}_{k,l}^*)$,  has been determined. More specifically, let $\mathcal{T}_{k,l}^*$ be $\bar{\mathcal{T}}_{k,l}^*$ shifted back to real time by performing inverses of the shifts, which were performed in previous iterations using (\ref{UpdateNan1}) and (\ref{UpdateNan2}). At any given time $t \in \mathcal{T}_{k,l}^*$, find the packet in $\mathcal{H}(\bar{\mathcal{T}}_{k,l}^*)$ that has arrived and has not finished transmission and has the \emph{earliest} upcoming deadline, transmit the said packet at rate $r(\bar{\mathcal{T}}_{k,l}^*)$. If no such packet can be found, then remain idle at time $t$.
For the packets whose transmission rate and schedule remains undetermined in this round of iteration, we update their arrival instant and deadline constraint according to (\ref{UpdateNan1}) and (\ref{UpdateNan2}), which basically says that the packets transmitted in interval $\mathcal{T}_{k,l}^*$ has been determined, and the transmission schedule for the remaining packets should be found by ignoring the time-interval $\mathcal{T}_{k,l}^*$, given that none of them is transmitted in $\mathcal{T}_{k,l}^*$. We iterate until the transmission rate and schedule of all the packets are found.

Algorithm 1 imply the following facts: the time-interval $\mathcal{T}_{k,l}^*$ is \emph{exclusively} used for the transmission of the packets in $\mathcal{H}(\bar{\mathcal{T}}_{k,l}^*)$. Moreover, no other time outside of $\mathcal{T}_{k,l}^*$ will be used for transmitting any packet from $\mathcal{H}(\bar{\mathcal{T}}_{k,l}^*)$. Furthermore, Step \ref{NanState04} of Algorithm 1 implies the following three points: first, the transmission schedule will not violate the causality constraint as it only transmits data upon its arrival. Second, due to the fact that it transmits data only upon its arrival, there may be idling periods during the interval $\mathcal{T}_{k,l}^*$. In fact, idling may occur if and only if all the packets that have arrived has finished transmission by time $t$, or all the packets that have arrived have a deadline earlier than time $t$. However, we shall proof in Theorem \ref{algorithm_optimal} that there is in fact no idling in the scheduler of Algorithm 1. Third, the deadline constraint is violated in the sense that if a packet has not finished transmission before its deadline, the remaining bits are never transmitted and we go on to transmit another packet with the next upcoming deadline given that it has already arrived.

Let $G$ be the total number of iterations that ran before Step \ref{NanState03} of Algorithm 1 is satisfied. The optimality and complexity of Algorithm 1 is described and proved in the following.

\begin{Lem}\label{Decreasing_Rate_Iter}
Assume $\mathcal{T}_{k,l}^{*g}$ is the sub-interval found in Step \ref{NanState02} of Algorithm \ref{Opt_Iter_Algorithm} in the $g$-th iteration, $g \in \{1,2,\cdots,G-1\}$, then $r(\mathcal{T}_{k,l}^{*g}) \ge r(\mathcal{T}_{k,l}^{*g+1})$.
\end{Lem}

Proof: Please see Appendix \ref{Proof_Decreasing_Rate_Iter}. \hfill $\Box$

\begin{Theo}\label{algorithm_optimal}
Algorithm \ref{Opt_Iter_Algorithm} is an optimal transmission schedule for the problem in (\ref{OptimizationPro}).
\end{Theo}

Proof: Please see Appendix \ref{Proof_algorithm_optimal}.  \hfill $\Box$

We now analyze the complexity of Algorithm 1. In each round of iteration, there are at most $N^2$ sub-intervals from the arrival instants of each packet to the deadline constraints of each packet. Thus, the complexity in each round of iteration is $\mathcal{O}(N^2)$. In addition, since there are $N$ packets in the packet sequence, and at each iteration, we determine the transmission schedule of at least one packet, the algorithm runs at most $G=N$ rounds of iterations. Thus, the complexity of the proposed algorithm is $\mathcal{O}(N^3)$.

\section{Conclusion}
We considered the problem of minimizing transmission energy consumption for packets with arbitrary arrival instants and deadline constraints over a point-to-point Additive White Gaussian Noise (AWGN) channel.
We first investigate the necessary and sufficient conditions of the optimal transmission scheduler. We then propose an algorithm which finds the transmission schedule of each packet in the order of the packets with the largest transmission rate to the packets with the smallest transmission rate. Finally, we show that our algorithm satisfies the sufficient conditions of the optimal transmission scheduler and thus, is optimal.

\appendices

\section{Proof of Lemma \ref{OptRateRelationship}} \label{Proof_OptRateRelationship}
 According to (\ref{Optimal_R}), we know that ${r_i^*}= g^{-1}( \beta _j^*  - \gamma _{i,j}^*)$, $j \in \mathcal{C}_i$, where $g^{-1}(\cdot)$ is monotonically increasing function, thus $r_i^*$ is monotonically increasing function with $\beta _j^*  - \gamma _{i,j}^*$.

According to Lemma \ref{NonIdlingtrans}, the optimal transmission in each epoch should be in ``non-idling'' mode, i.e.,  ${\sum\limits_{i \in \mathcal{F}_j } {{\tau _{i,j}^*} - \left| {{\mathcal{E}_j}} \right|}  = 0}$, $j=1,\cdots,M$. This means that $\Psi_j$ is non-empty, and
%
 %
$\sum\limits_{i \in \Psi_j}{\tau_{i,j}^*} = |\mathcal{E}_j|$ holds. According to (\ref{ComplementarySlackness}b), we see that $\gamma _{i,j}^*=0$ for all $i \in \Psi_j$. Substituting this into (\ref{Optimal_R}) gives ${r_i^*}= g^{-1}( \beta _j^*)$, $\forall i \in \Psi_j$, which means that all packets that are in $\Psi_j$ are transmitted with the same rate.


Meanwhile, since $\tau_{k,j}^* = 0$ for all $k \in \bar{\Psi}_j$, according to (\ref{ComplementarySlackness}b), we see that $\gamma _{k,j}^* \geq 0$ for all $k \in \bar{\Psi}_j$. Substituting this into (\ref{Optimal_R}) gives $r_k^* = g^{-1}( \beta _j^*  - \gamma _{k,j}^*)$, $\forall k \in \bar{\Psi}_j$. Since $g^{-1}(\cdot)$ is a monotonically increasing function, we have $r_i^* = g^{-1}( \beta _j^*) \ge g^{-1}( \beta _j^*  - \gamma _{k,j}^*)= r_k^*$, $\forall i \in \Psi_j, \forall k \in \bar{\Psi}_j$.
%
%
%
%
\hfill $\Box$

\section{Proof of Theorem \ref{NS Conditions}} \label{Proof_NS Conditions}
%
%
We prove by contradiction.
Assume that there exists a feasible scheduler $\mathcal{S}^N$ which satisfies all the conditions of Lemmas \ref{ConstantRate}, \ref{NonIdlingtrans} and \ref{OptRateRelationship} but is not the optimal transmission scheduler.
Meanwhile, the optimal transmission scheduler is denoted as $\mathcal{S}^O$ which, according to the necessary conditions, satisfies all the conditions in Lemmas \ref{ConstantRate}, \ref{NonIdlingtrans} and \ref{OptRateRelationship}. Here, the superscript $N$ and $O$ represent ``Non-Optimal'' and ``Optimal'', respectively. According to the the definition of $\mathcal{S}^N$ and $\mathcal{S}^O$, we have $E^N >E^O$, where $E^N$ ($E^O$) is the transmission energy of scheduler $\mathcal{S}^N$ ($\mathcal{S}^O$). This implies, by the power-rate relationship function (\ref{PowerRate}), that at least one packet in $\mathcal{S}^N$ has a larger transmission rate than that of the same packet in $\mathcal{S}^O$, i.e., there exists an $\bar{i} \in \{1,2, \cdots, N\}$ such that $r_{\bar{i}}^N > r_{\bar{i}}^O$.
We denote by $\Omega$ as the set of packets satisfying $r_i^N > r_i^O, \{1,2, \cdots, N\}$, i.e., $\Omega=\{i|r_i^N > r_i^O\}$, and we know that $\Omega \neq \phi$.  Based on the definition of $\Omega$, it is obvious that it takes less time in $\mathcal{S}^N$ than $\mathcal{S}^O$ to complete the transmission of all the packets in set $\Omega$, i.e.,
\begin{align}\label{lesstime}
  \sum\limits_{i \in \Omega}\sum\limits_{j \in \mathcal{C}_i}{\tau_{i,j}^N} < \sum\limits_{i \in \Omega}\sum\limits_{j \in \mathcal{C}_i}{\tau_{i,j}^O}
\end{align}
which means that there must exist at least one epoch $\mathcal{E}_{\bar{j}}$, where $\bar{j} \in \mathcal{C}_i$ for some $i \in \Omega$, that satisfies
\begin{align}\label{lesstimesingleepoch}
  \sum\limits_{i \in \Omega} {\tau_{i,\bar{j}}^N} <  \sum\limits_{i \in \Omega} {\tau_{i,\bar{j}}^O}
\end{align}
This implies that there exists a packet $\bar{k} \in \Omega \bigcap \mathcal{F}_{\bar{j}}$, where $\tau_{\bar{k},\bar{j}}^O>\tau_{\bar{k},\bar{j}}^N \geq 0$.
Concentrating on epoch $\bar{j}$, according to Lemma \ref{NonIdlingtrans}, we have $\sum\limits_{i \in \mathcal{F}_{\bar{j}}} {\tau_{i,\bar{j}}^N} = \sum\limits_{i \in \mathcal{F}_{\bar{j}}}{\tau_{i,\bar{j}}^O}= |\mathcal{E}_{\bar{j}}|$, so there must exist another packet $\bar{l}$, $\bar{l} \in \mathcal{F}_{\bar{j}}\backslash \Omega$, such that $\tau_{\bar{l},\bar{j}}^N > \tau_{\bar{l},\bar{j}}^O \geq 0$. So for scheduler $\mathcal{S}^O$, we have $\tau_{\bar{k},\bar{j}}^O> 0$ and $\tau_{\bar{l},\bar{j}}^O \geq 0$. According to Lemma \ref{OptRateRelationship}, we have $r_{\bar{k}}^O \geq r_{\bar{l}}^O$. Similarly, for scheduler $\mathcal{S}^N$, we have $r_{\bar{l}}^N \geq r_{\bar{k}}^N$. Based on the fact that $\bar{k} \in \Omega$, we have $r_{\bar{l}}^N \geq r_{\bar{k}}^N>r_{\bar{k}}^O \geq r_{\bar{l}}^O$, which means that $\bar{l} \in \Omega$, and this contradicts the assumption $\bar{l} \in \mathcal{F}_{\bar{j}}\backslash \Omega$. This contradiction illustrates that all conditions in Lemma \ref{ConstantRate}, \ref{NonIdlingtrans} and \ref{OptRateRelationship} are sufficient conditions for the optimality of the scheduler.
 \hfill $\Box$

\section{Proof of Lemma \ref{Decreasing_Rate_Iter}} \label{Proof_Decreasing_Rate_Iter}
We prove by contradiction. Note that we choose to write the proof using the original arrival instants and deadline constraints of the packets, but the same argument follows if we use the updated arrival instants and deadline constraints, since they are simply shifted versions of each other.
We assume that $r(\mathcal{T}_{k,l}^{*g}) < r(\mathcal{T}_{k,l}^{*g+1})$, i.e., the maximum transmission rate selected in the $g$-th round of iteration is strictly less than that of the $(g+1)$-th round of iteration. Denote $\mathcal{T}_{k,l}^{*g}=[t_{a,k}^{*g}, t_{d,l}^{*g}]$. There are two possibilities to consider.
\begin{enumerate}
 \item The sub-interval $\mathcal{T}_{k,l}^{*g+1} \subseteq [0,t_{a,k}^{*g})$ or $\mathcal{T}_{k,l}^{*g+1} \subseteq (t_{d,l}^{*g}, T]$. So both sub-intervals $\mathcal{T}_{k,l}^{*g}$ and $\mathcal{T}_{k,l}^{*g+1}$ will be considered in the $g$-th round of iteration in Step \ref{NanState} of Algorithm 1. Furthermore,
 %
 %
$r(\mathcal{T}_{k,l}^{*g+1})$ does not change before and after the removing of $\mathcal{T}_{k,l}^{*g}$ since the removing of $\mathcal{T}_{k,l}^{*g}$ does not result in any change of the packets set with $\mathcal{L}_i \subseteq \mathcal{T}_{k,l}^{*g+1}$. Thus, the assumption $r(\mathcal{T}_{k,l}^{*g}) < r(\mathcal{T}_{k,l}^{*g+1})$, is \emph{contradicting} the fact that Algorithm \ref{Opt_Iter_Algorithm} selected the sub-interval with the maximum transmission rate in the $g$-th round of iteration.
\item The sub-interval $\mathcal{T}_{k,l}^{*g+1}$ does not satisfy the condition of the previous sub-case, i.e., $\mathcal{T}_{k,l}^{*g+1}$ either contains time point $t_{a,k}^{*g}$ or $t_{d,l}^{*g}$ or both. This means that in the $g$-th round of iteration, both sub-intervals $\mathcal{T}_{k,l}^{*g}$ and $\mathcal{T}_{k,l}^{*g} \bigcup \mathcal{T}_{k,l}^{*g+1}$ will be considered in Step \ref{NanState} of Algorithm 1. Let $\mathcal{H}^g(\mathcal{T}_{k,l})$ denote the set of packets whose life time duration is contained in sub-interval $\mathcal{T}_{k,l}$ in the $g$-th round of iteration. Then, according to Algorithm 1, we have
\begin{align}
\mathcal{H}^g (\mathcal{T}_{k,l}^{*g} \cup \mathcal{T}_{k,l}^{*g+1})=\mathcal{H}^g(\mathcal{T}_{k,l}^{*g}) \cup \mathcal{H}^{g+1}(\mathcal{T}_{k,l}^{*g+1})
\end{align}
This means that the rate of $\mathcal{T}_{k,l}^{*g} \cup \mathcal{T}_{k,l}^{*g+1}$ computed in the $g$-th round of iteration is
 \begin{align}\label{Unionrate}
r(\mathcal{T}_{k,l}^{*g} \cup \mathcal{T}_{k,l}^{*g+1})   = \frac {\sum\limits_{i \in \mathcal{H}^g(\mathcal{T}_{k,l}^{*g})} {B_i}  +  \sum\limits_{i \in \mathcal{H}^{g+1}(\mathcal{T}_{k,l}^{*g+1} )} {B_i} }  {|\mathcal{T}_{k,l}^{*g}|  + |\mathcal{T}_{k,l}^{*g+1}| }
 \end{align}
Due to assumption of $r(\mathcal{T}_{k,l}^{*g}) < r(\mathcal{T}_{k,l}^{*g+1})$, we have
 \begin{align}\label{InequalityAssumption}
   r (\mathcal{T}_{k,l}^{*g}) &= \frac {\sum\limits_{i \in \mathcal{H}^g(\mathcal{T}_{k,l}^{*g})} {B_i}} {|\mathcal{T}_{k,l}^{*g}|}\nonumber \\
   & < \frac {\sum\limits_{i\in \mathcal{H}^{g+1}(\mathcal{T}_{k,l}^{*g+1} )} {B_i}} {|\mathcal{T}_{k,l}^{*g+1}|} = r (\mathcal{T}_{k,l}^{*g+1})
 \end{align}
which implies
\begin{align}\label{}
   r (\mathcal{T}_{k,l}^{*g}) &= \frac {\sum\limits_{i \in \mathcal{H}^g(\mathcal{T}_{k,l}^{*g})} {B_i}} {|\mathcal{T}_{k,l}^{*g}|}\nonumber \\
  & < \frac {\sum\limits_{i \in \mathcal{H}^g(\mathcal{T}_{k,l}^{*g})} {B_i}  +  \sum\limits_{i \in \mathcal{H}^{g+1}(\mathcal{T}_{k,l}^{*g+1} )} {B_i} }  {|\mathcal{T}_{k,l}^{*g}|  + |\mathcal{T}_{k,l}^{*g+1}| }=r(\mathcal{T}_{k,l}^{*g} \cup \mathcal{T}_{k,l}^{*g+1}) \nonumber
 \end{align}
Since in the $g$-th round of iteration, both sub-intervals $\mathcal{T}_{k,l}^{*g}$ and $\mathcal{T}_{k,l}^{*g} \bigcup \mathcal{T}_{k,l}^{*g+1}$ will be considered in Step \ref{NanState} of Algorithm 1, this contradicts with the fact that Algorithm \ref{Opt_Iter_Algorithm} selectes the sub-interval with the maximum transmission rate in the $g$-th round of iteration.

 \end{enumerate}
Thus, we have shown a contradiction in each of the two possible cases, which means that the assumption $r(\mathcal{T}_{k,l}^{*g}) < r(\mathcal{T}_{k,l}^{*g+1})$ does not hold, and we in fact have $r(\mathcal{T}_{k,l}^{*g}) \geq r(\mathcal{T}_{k,l}^{*g+1})$ for all $g \in \{1,2,\cdots, G-1\}$. \hfill $\Box$

\section{Proof of Theorem \ref{algorithm_optimal}} \label{Proof_algorithm_optimal}
To prove the optimality of the scheduler in Algorithm 1, we invoke Theorem \ref{NS Conditions}, i.e., we will prove that Algorithm 1 provides a feasible scheduler, and also, the scheduler of Algorithm 1 satisfies Lemmas \ref{ConstantRate},  \ref{NonIdlingtrans} and \ref{OptRateRelationship}.

First, from the description of Algorithm 1, it is easy to see that all packets in a selected sub-interval, i.e.,  $\mathcal{H}(\mathcal{T}_{k,l}^{*g})$ of round $g$, $g \in \{1,2,\cdots,G\}$, are transmitted with the same rate. Thus, each packet is transmitted with a constant rate and Lemma \ref{ConstantRate} is satisfied.

Next, we prove that there are no idling periods using the scheduler in Algorithm 1 by contradiction. We choose to write the proof using the original arrival instants and deadline constraints of the packets, but the same argument follows if we use the updated arrival instants and deadline constraints, since they are simply shifted versions of each other. Suppose at round $g$, $g \in \{1,2,\cdots,G\}$, there are idling periods in
$\mathcal{T}_{k,l}^{*g}=[t_{a,k}^{*g}, t_{d,l}^{*g}]$, and we denote the first idling period as $[t_1^g, t_2^g] \subseteq [t_{a,k}^{*g}, t_{d,l}^{*g}]$. There are the following three cases:
\begin{enumerate}
\item $t_1^g=t_{a,k}^{*g}$: in this case, packet $k$ has arrived but it is not being transmitted because $k \notin \mathcal{H}(\bar{\mathcal{T}}_{k,l}^{*g})$. Let the earliest arrival instants of the packets in $\mathcal{H}(\bar{\mathcal{T}}_{k,l}^{*g})$ be $t_{a,\bar{k}}^{g}$, and we have $t_{a,\bar{k}}^{g} > t_{a,k}^{*g}$, then, the rate of the sub-interval $[t_{a,\bar{k}}^{g}, t_{d,l}^{*g}]$ is
\begin{align}
\frac{\sum\limits_{i \in \mathcal{H}(\bar{\mathcal{T}}_{k,l}^{*g})} B_i}{t_{d,l}^{*g}-t_{a,\bar{k}}^{g}}
\end{align}
which is strictly larger than $r(\bar{\mathcal{T}}_{k,l}^{*g})$ which is equal to
\begin{align}
\frac{\sum\limits_{i \in \mathcal{H}(\bar{\mathcal{T}}_{k,l}^{*g})} B_i}{t_{d,l}^{*g}-t_{a,k}^{*g}}
\end{align}
due to the fact that $t_{a,\bar{k}}^{g} > t_{a,k}^{*g}$. This contradicts with Algorithm 1 where $r(\bar{\mathcal{T}}_{k,l}^{*g})$ is the sub-interval with the largest transmission rate in round $g$.
\item $t_1^g > t_{a,k}^{*g}$ and $t_2^g<t_{d,l}^{*g}$: in this case, idling happens because all packets in $\mathcal{H}(\bar{\mathcal{T}}_{k,l}^{*g})$ that arrived before $t_1^g$ have either finished transmitting by $t_1^g$ or have a deadline earlier than $t_1^g$. By the idling policy of Algorithm 1, there are no packets arriving in the period of $[t_1^g, t_2^g]$, and $t_2^g$ is the arrival instant of some packet in $\mathcal{H}(\bar{\mathcal{T}}_{k,l}^{*g})$, thus ending the idling period. Thus, $[t_2^g, \bar{t}_{d,l}^{*g}]$ is a sub-interval.
We have two sub-cases:
\begin{enumerate}
\item Sub-case 1: consider the packets in $\mathcal{H}(\bar{\mathcal{T}}_{k,l}^{*g})$ that arrived before $t_1^g$, all of them have finished transmitting before $t_1^g$. In this case, denote $\mathcal{U}_1^g$ as the set of packets that are in $\mathcal{H}(\bar{\mathcal{T}}_{k,l}^{*g})$ and has arrival instant before $t_2^g$, i.e., $\mathcal{U}_1^g=\{i|i \in \mathcal{H}(\bar{\mathcal{T}}_{k,l}^{*g}), t_{a,i} < t_{2}^g\}$. Then, the set of packets whose life time duration is contained in sub-interval $[t_2^g, t_{d,l}^{*g}]$ is
$\mathcal{H}(\bar{\mathcal{T}}_{k,l}^{*g}) \backslash \mathcal{U}_1^g$.
The transmission rate of the packets in $\mathcal{U}_1^g$ is
\begin{align}
\frac{\sum\limits_{i \in \mathcal{U}_1^g} B_i   }{t_1^g-t_{a,k}^{*g}} \label{NanDetails01}
\end{align}
where the numerator is because in this sub-case, all data of packets in $\mathcal{U}_1^g$ have finished transmission, and the denominator is
due to the assumption that $[t_1^g, t_2^g]$ is the \emph{first} idling period in $\mathcal{T}_{k,l}^{*g}$, and therefore, the packets in $\mathcal{U}_1^g$ are transmitted from $t_{a,k}^{*g}$ to $t_1^g$ continuously. According to Algorithm 1, this rate is equal to $r(\bar{\mathcal{T}}_{k,l}^{*g})$ which is equal to
\begin{align}
\frac{\sum\limits_{i \in \mathcal{U}_1^g} B_i+ \sum\limits_{i \in \mathcal{H}(\bar{\mathcal{T}}_{k,l}^{*g}) \backslash \mathcal{U}_1^g} B_i}{(t_1^g-t_{a,k}^{*g})+(t_2^g-t_1^g)+(t_{d,l}^{*g}-t_2^g)} \label{NanDetails02}
\end{align}
Equating (\ref{NanDetails01}) and (\ref{NanDetails02}), we have
\begin{align}
r(\bar{\mathcal{T}}_{k,l}^{*g})=\frac{\sum\limits_{i \in \mathcal{U}_1^g} B_i   }{t_1^g-t_{a,k}^{*g}} = \frac{ \sum\limits_{i \in \mathcal{H}(\bar{\mathcal{T}}_{k,l}^{*g}) \backslash \mathcal{U}_1^g} B_i}{(t_2^g-t_1^g)+(t_{d,l}^{*g}-t_2^g)}
\end{align}
which is strictly smaller than
\begin{align}
 \frac{ \sum\limits_{i:\mathcal{H}(\bar{\mathcal{T}}_{k,l}^{*g}) \backslash \mathcal{U}_1^g} B_i}{t_{d,l}^{*g}-t_2^g}
\end{align}
which is the rate of the sub-interval $[t_2^g, t_{d,l}^{*g}]$. This \emph{contradicts} with the fact that $r(\bar{\mathcal{T}}_{k,l}^{*g})$ is the sub-interval with the largest transmission rate in round $g$.
\item Sub-case 2: consider the packets in $\mathcal{H}(\bar{\mathcal{T}}_{k,l}^{*g})$ that have arrived before $t_1^g$, at least one of these packets did not finish transmitting and was cut off because it had reached its deadline before completion. Among all the packets that have not finished transmitting before their deadline, let $\bar{j}$ be the packet with the earliest deadline. Suppose the amount of time for the transmission of Packet $\bar{j}$ is $t_{\bar{j}}$ using Algorithm 1, since Packet $\bar{j}$ is unfinished before its deadline, we know
\begin{align}
t_{\bar{j}} & <\frac{B_{\bar{j}}}{r(\bar{\mathcal{T}}_{k,l}^{*g})}
\end{align}
On the other hand, consider the sub-interval $\mathcal{I} \triangleq [t_{a,\bar{j}},t_{d,\bar{j}}]$. , we have
\begin{align}
t_{\bar{j}} & \leq |\mathcal{I}|-\sum_{i \in \mathcal{H}(\mathcal{I}) \backslash \{\bar{j}\}} \frac{B_i}{r(\bar{\mathcal{T}}_{k,l}^{*g})} \label{NanProof01}
\end{align}
where (\ref{NanProof01}) follows because Packet $\bar{j}$ is the first packet to be unfinished by its deadline, it means that all other packets in $\mathcal{H}(\mathcal{I})$ have finished transmission, i.e., has been transmitted for the time of $\frac{B_i}{r(\bar{\mathcal{T}}_{k,l}^{*g})}$, $i \in \mathcal{H}(\mathcal{I}) \backslash \{\bar{j}\}$. We have inequality rather than equality because Algorithm 1 could have used the sub-interval $\mathcal{I}$ to transmit some packets who are not $\mathcal{H}(\mathcal{I})$.
Thus, we have the following 2 cases:
\begin{enumerate}
\item In case 1, we have
\begin{align}
|\mathcal{I}|-\sum_{i \in \mathcal{H}(\mathcal{I}) \backslash \{\bar{j}\}} \frac{B_i}{r(\bar{\mathcal{T}}_{k,l}^{*g})}
<\frac{B_{\bar{j}}}{r(\bar{\mathcal{T}}_{k,l}^{*g})} \label{NanProof03}
\end{align}
This means
\begin{align}
t_{\bar{j}} & \leq |\mathcal{I}|-\sum_{i \in \mathcal{H}(\mathcal{I}) \backslash \{\bar{j}\}} \frac{B_i}{r(\bar{\mathcal{T}}_{k,l}^{*g})}  \nonumber\\
&<\frac{B_{\bar{j}}}{r(\bar{\mathcal{T}}_{k,l}^{*g})}
\end{align}
which means
\begin{align}
 |\mathcal{I}| <  \frac{ \sum_{i \in \mathcal{H}(\mathcal{I}) } B_i}{r(\bar{\mathcal{T}}_{k,l}^{*g})}
\end{align}
which further means that
\begin{align}
r(\mathcal{I})=\frac{ \sum_{i \in \mathcal{H}(\mathcal{I}) } B_i}{|\mathcal{I}|}>r(\bar{\mathcal{T}}_{k,l}^{*g}) \label{NanProof02}
\end{align}
However, (\ref{NanProof02}) is a \emph{contradiction} to the fact that $r(\bar{\mathcal{T}}_{k,l}^{*g})$ is the largest rate among all sub-intervals in round $g$.
\item (\ref{NanProof03}) is not true, i.e.,
\begin{align}
|\mathcal{I}|-\sum_{i \in \mathcal{H}(\mathcal{I}) \backslash \{\bar{j}\}} \frac{B_i}{r(\bar{\mathcal{T}}_{k,l}^{*g})}
 \geq \frac{B_{\bar{j}}}{r(\bar{\mathcal{T}}_{k,l}^{*g})} \label{NanProof04}
\end{align}
In this case, if Algorithm 1 had used sub-interval $\mathcal{I}$ to transmit only the packets in $\mathcal{H}(\mathcal{I})$, then Packet $\bar{j}$ would have finished transmitting. The reason why Packet $\bar{j}$ has not finished transmitting is because Algorithm 1 has used some time in sub-interval $\mathcal{I}$ to transmit some packets that are not in $\mathcal{H}(\mathcal{I})$. Denote the set of such packets as $\mathcal{K}(\mathcal{I})$. According to Step \ref{NanState04} of Algorithm 1, a packet, that is not in $\mathcal{H}(\mathcal{I})$, would only be transmitted during the period of $\mathcal{I}$ if it had an arrival instant earlier $t_{a,\bar{j}}$ and a deadline constraint in $[t_{a,\bar{j}},t_{d,\bar{j}}]$. Note that a packet with a later deadline than $t_{d,\bar{j}}$ would not be transmitted in $[t_{a,\bar{j}},t_{d,\bar{j}}]$ because Packet $\bar{j}$ has already arrived and since it did not finish transmission by its deadline, it would not leave any window of time open in $[t_{a,\bar{j}},t_{d,\bar{j}}]$ for the transmission of a packet with a later deadline. Let Packet $\bar{k}$ be the packet in $\mathcal{K}(\mathcal{I})$ with the earliest arrival instant, i.e., $\bar{k}=\arg \min\limits_{i \in \mathcal{K}(\mathcal{I})} t_{a,i}$.

Now, consider the sub-interval $[t_{a,\bar{k}}, t_{d,\bar{j}}]$. We again have the two cases as described in (\ref{NanProof03}) and (\ref{NanProof04}) where the interval $\mathcal{I}$ is redefined as $\mathcal{I} \triangleq [t_{a,\bar{k}}, t_{d,\bar{j}}]$. In the case that (\ref{NanProof03}) is true, we again arrive at the contradiction with the fact that $r(\bar{\mathcal{T}}_{k,l}^{*g})$ is the largest rate among all sub-intervals in round $g$. In the case that (\ref{NanProof04}) is true, we conclude that again that the reason why Packet $\bar{j}$ has not finished transmitting is because Algorithm 1 has used some time in sub-interval $\mathcal{I}$ to transmit some packets that are not in $\mathcal{H}(\mathcal{I})$. Now, we analyze what kind of packets would be transmitted in $\mathcal{I}$ and not be in $\mathcal{H}(\mathcal{I})$ for $\mathcal{I} = [t_{a,\bar{k}}, t_{d,\bar{j}}]$.
According to Step \ref{NanState04} of Algorithm 1, a packet that has an arrival instant earlier than $t_{a,\bar{k}}$ and a deadline constraint in $[t_{a,\bar{k}}, t_{d,\bar{j}}]$ could be transmitted in $[t_{a,\bar{k}}, t_{d,\bar{j}}]$. It would seem that, since in this case the starting point of the interval is $t_{a,\bar{k}}$ which satisfies $t_{a,\bar{k}} < t_{a,\bar{j}}$, that a packet with a deadline later than $t_{d,\bar{j}}$ and an arrival instant earlier than $t_{a,\bar{j}}$ could possibly be transmitted due to the fact that the packets in $\mathcal{H}([t_{a,\bar{j}}, t_{d,\bar{j}}])$ have not arrived before $t_{a,\bar{j}}$. However, this is not true because Packet $\bar{k}$ in $\mathcal{K}([t_{a,\bar{j}}, t_{d,\bar{j}}])$ have transmitted into the sub-interval $[t_{a,\bar{j}}, t_{d,\bar{j}}]$, which leaves no window of time open in $[t_{a,\bar{k}}, t_{a,\bar{j}}]$ for the transmission of a packet with a later deadline, i.e., whenever there is time in the interval of $[t_{a,\bar{k}}, t_{a,\bar{j}}]$, rather than scheduling a packet with a later deadline than $t_{d,\bar{j}}$, Packet $\bar{k}$ would be transmitting. As for the time of $[t_{a,\bar{j}}, t_{d, \bar{j}}]$, rather than scheduling a packet with a later deadline than $t_{d,\bar{j}}$, Packet $\bar{j}$ would be transmitting. So the only packets transmitted in $[t_{a,\bar{k}}, t_{d,\bar{j}}]$ but not in $\mathcal{H}([t_{a,\bar{k}}, t_{d,\bar{j}}])$ are packets that have an arrival instant earlier than $t_{a,\bar{k}}$ and a deadline constraint in $[t_{a,\bar{k}}, t_{d,\bar{j}}]$. Denote the set of packets again by $\mathcal{K}(\mathcal{I})$, where $\mathcal{I}=[t_{a,\bar{k}}, t_{d,\bar{j}}]$. And further let Packet $\tilde{k}$ be the packet in $\mathcal{K}(\mathcal{I})$ with the earliest arrival instant, i.e., $\tilde{k}=\arg \min\limits_{i \in \mathcal{K}(\mathcal{I})} t_{a,i}$.

    Now consider the sub-interval $[t_{a,\tilde{k}}, t_{d,\bar{j}}]$. This case follows the case of the sub-interval $[t_{a,\bar{k}}, t_{d,\bar{j}}]$ exactly with $\mathcal{I}=[t_{a,\tilde{k}}, t_{d,\bar{j}}]$. We have again the two cases as described in (\ref{NanProof03}) and (\ref{NanProof04}) and we either arrive at a contradiction or we enlarge the sub-interval to $[t_{a,\hat{k}}, t_{d,\bar{j}}]$. We iterate until we either arrive a contradiction at some step, or we have enlarged the interval to $[t_{a,k}, t_{d,\bar{j}}]$, where $t_{a,k}$ is the starting point of $\mathcal{T}_{k,l}^{*g}$.  In this sub-interval, we would not have the case described in (\ref{NanProof04}) because according to Algorithm 1, there are no packets with an earlier arrival instant than $t_{a,k}$ transmitted in $\mathcal{T}_{k,l}^{*g}$. So we are left with the case described in (\ref{NanProof03}) only, and we arrive at a \emph{contradiction}. Hence, we will always get a \emph{contradiction} for Sub-case 2.
\end{enumerate}
\end{enumerate}

\item $t_1^g > t_{a,k}^{*g}$ and $t_2^g=t_{d,l}^{*g}$; in this case, by the idling policy of Algorithm 1, there are no packets arriving in the period of $[t_1^g, t_{d,l}^{*g}]$. All the packets in $\mathcal{H}(\bar{\mathcal{T}}_{k,l}^{*g})$ arrived before $t_1^g$.
Similar to the previous case, we have two sub-cases:
\begin{enumerate}
\item Sub-case 1: all the packets in $\mathcal{H}(\bar{\mathcal{T}}_{k,l}^{*g})$ have finished transmitting before $t_1^g$, where $t_1^g<t_{d,l}^{*g}$, but this is not possible since this would mean that the data are transmitted at the rate of
\begin{align}
\frac{\sum\limits_{i: \mathcal{H}(\bar{\mathcal{T}}_{k,l}^{*g})} B_i   }{t_1^g-t_{a,k}^{*g}}
\end{align}
which is strictly larger than the actual rate of transmission $r(\bar{\mathcal{T}}_{k,l}^{*g})=\frac{\sum\limits_{i: \mathcal{H}(\bar{\mathcal{T}}_{k,l}^{*g})} B_i   }{t_2^g-t_{a,k}^{*g}}$. Thus, we have a \emph{contradiction}.
\item Sub-case 2: consider the packets in $\mathcal{H}(\bar{\mathcal{T}}_{k,l}^{*g})$, at least one of these packets did not finish transmitting and was cut off because it had reached its deadline before completion. This sub-case is exactly the same as the sub-case 2 of $t_1^g > t_{a,k}^{*g}$ and $t_2^g<t_{d,l}^{*g}$, where the arguments do not depend on whether $t_2^g=t_{d,l}^{*g}$ or not. Thus, we have a \emph{contradiction} for this sub-case too.
\end{enumerate}
\end{enumerate}
Since we have a contradiction for each of the above three cases, we have proven that Algorithm 1 does not generate a scheduler with idling periods. Thus, it satisfies Lemma 2.

Thirdly, we prove the feasibility of the scheduler generated by Algorithm 1. Based on Step \ref{NanState04} in Algorithm 1, we only transmit data upon its arrival. So the scheduler in Algorithm 1 always satisfy the causality constraint. We now prove that it satisfies the deadline constraint as well. Based on Step \ref{NanState04} of Algorithm 1, it violates the deadline constraint only when there exists some packet whose data is not completely transmitted before its deadline. But all data is transmitted at the minimum transmission rate of $r(\bar{\mathcal{T}}_{k,l}^{*g})$ in each round $g$. If some packet is not completely transmitted, then, we would have transmitted less data than $\sum\limits_{i \in \mathcal{H}(\mathcal{T}_{k,l}^{*g})} B_i$ and there would be some idling period. Since we have already proved that Algorithm 1 does not have any idling period, it means that all the data is completely transmitted by its deadline and Algorithm 1 satisfies the deadline constraint as well. Thus, Algorithm 1 is feasible.

Finally, we prove that Algorithm \ref{Opt_Iter_Algorithm} satisfies Lemma \ref{OptRateRelationship}. Since there exists no epoch who belongs to two sub-intervals, and we know that each sub-interval is transmitted with the equal rate of $r(\mathcal{T}_{k,l}^{*g})$, it means that each epoch is also transmitted with the same rate. Thus, proving (1) of Lemma \ref{OptRateRelationship}. Consider an epoch $\mathcal{E}_j \in \mathcal{T}_{k,l}^{*g}$, for some $g$ in $\{1,2,\cdots, G\}$, consider all
packets with a higher transmission rate than $r(\mathcal{T}_{k,l}^{*g})$, then  according to Lemma \ref{Decreasing_Rate_Iter}, they must have been determined to transmit in Iteration $1$ or $2$ or $\cdots$ or $g-1$. This means that the life time of such packets are contained in $\mathcal{T}_{k,l}^{*m}$ for some $m \in \{1,2,\cdots,g-1\}$, and correspondingly, they are not in $\mathcal{F}_j$. This means that for all packets in $\mathcal{F}_j$, their transmission rate can not be larger than $r(\mathcal{T}_{k,l}^{*g})$, thus proving (2) of Lemma \ref{OptRateRelationship}. To sum up,  Algorithm \ref{Opt_Iter_Algorithm} satisfies Lemma \ref{OptRateRelationship}.

Since all the conditions in Lemmas \ref{ConstantRate}, \ref{NonIdlingtrans} and \ref{OptRateRelationship} are sufficient conditions of optimality for the problem in (\ref{OptimizationPro}), we have proved that Algorithm \ref{Opt_Iter_Algorithm} indeed finds the optimal transmission schedule.

\bibliographystyle{IEEEtran}
\bibliography{Myreference}

\end{document}